\documentclass[%
reprint,
superscriptaddress,
 aps,
]{revtex4-2}

\usepackage{graphicx}
\usepackage{dcolumn}
\usepackage{bm}
\usepackage{amssymb}

\usepackage{hyperref}
\hypersetup{colorlinks=true, linkcolor=blue, urlcolor=black, filecolor=magenta, citecolor=blue}
\usepackage{float}

\begin{document}

\preprint{APS/123-QED}

\title{Compact superconducting microwave resonators based on Al-AlO$_x$-Al capacitors}

\author{Julia Zotova} 
\email{yuliya.zotova@phystech.edu}
\affiliation{Skolkovo Institute of Science and Technology, 121205 Moscow, Russia}
\affiliation{Moscow Institute of Physics and Technology, Institutskiy Pereulok 9, Dolgoprudny 141701, Russia}
\affiliation{RIKEN Center for Quantum Computing (RQC), Wako, Saitama 351-0198, Japan}

\author{Rui Wang}%
\affiliation{Department of Physics, Tokyo University of Science, 1–3 Kagurazaka, Shinjuku, Tokyo 162–0825, Japan}
\affiliation{RIKEN Center for Quantum Computing (RQC), Wako, Saitama 351-0198, Japan}

\author{Alexander Semenov}%
\affiliation{Moscow Institute of Physics and Technology, Institutskiy Pereulok 9, Dolgoprudny 141701, Russia}
\affiliation{Moscow State Pedagogical University, Malaya Pirogovskaya Street 1/1, Moscow 119435, Russia}

\author{Yu Zhou}%
\affiliation{RIKEN Center for Quantum Computing (RQC), Wako, Saitama 351-0198, Japan}

\author{Ivan Khrapach}%
\affiliation{Moscow Institute of Physics and Technology, Institutskiy Pereulok 9, Dolgoprudny 141701, Russia}
\affiliation{Russian Quantum Center, 121205 Skolkovo, Russia}

\author{Akiyoshi Tomonaga}%
\affiliation{Department of Physics, Tokyo University of Science, 1–3 Kagurazaka, Shinjuku, Tokyo 162–0825, Japan}
\affiliation{RIKEN Center for Quantum Computing (RQC), Wako, Saitama 351-0198, Japan}

\author{Oleg Astafiev}%
\affiliation{Skolkovo Institute of Science and Technology, 121205 Moscow, Russia}
\affiliation{Moscow Institute of Physics and Technology, Institutskiy Pereulok 9, Dolgoprudny 141701, Russia}

\author{Jaw-Shen Tsai}%
\email{tsai@riken.jp}
\affiliation{Department of Physics, Tokyo University of Science, 1–3 Kagurazaka, Shinjuku, Tokyo 162–0825, Japan}
\affiliation{RIKEN Center for Quantum Computing (RQC), Wako, Saitama 351-0198, Japan}

\date{\today}

\begin{abstract}
We address the scaling-up problem for superconducting quantum circuits by using lumped-element resonators based on an alternative fabrication method of aluminum -- aluminum oxide -- aluminum  ($\mathrm{Al/AlO_x/Al}$) parallel-plate capacitors. The size of the resonators is only 0.04~$\mathrm{mm^2}$, which is more than one order smaller than the typical size of coplanar resonators (1~$\mathrm{mm^2}$). The fabrication method we developed easily fits into the standard superconducting qubits fabrication process.  We have obtained capacitance per area 14~fF/$\mathrm{\mu m^2}$ and the internal quality factor $\mathrm{1\times 10^3 - 8\times 10^3}$ at the single-photon level.  Our results show that such devices based on $\mathrm{Al/AlO_x/Al}$ capacitors could be further applied to the qubit readout scheme, including resonators, filters, amplifiers, as well as microwave metamaterials and innovative types of qubits, such as $0-\pi$ qubit.
\end{abstract}

\maketitle

\section{\label{sec:introduction}Introduction}
Microwave components, such as resonators, transmission lines, filters and amplifiers, play key roles in superconducting quantum circuits~\cite{krantz2019quantum}. In particular,   coplanar waveguide (CPW) structures are widely used because of their simplicity and high quality factors. Inductance and capacitance of such resonators are distributed, and frequency is determined by their length~\cite{pozar2011microwave}.  As a result, CPW resonators have a relatively large size (1~$\mathrm{mm^2}$) in the operating frequency range of several gigahertz (GHz).  Hence, CPW resonators are sensitive to the inhomogeneous ground potential problem, which may cause slotline modes; to suppress these parasitic modes additional care is required~\cite{chen2014fabrication}. Also, coplanar waveguide resonators have a limitation of coupling strength and the characteristic impedance.  Due to a coplanar waveguide using a substrate as a dielectric, the electric field distribution can cause parasitic crosstalks. Also, a large size of CPW resonators could be a problem for the global trend of scaling up quantum systems. 

As an alternative to distributed-element structures, lumped-element resonators could be used. Their frequency is determined by only values of inductance and capacitance and is independent of their geometric size. Achieving 50~$\Omega$ resonator impedance in the gigahertz range requires relatively large capacitance of a few hundreds of femtofarads up to one picofarad. To make such capacitors on a chip a planar structure -- such as an interdigital capacitor (IDC) -- is generally used~\cite{geerlings2012improving}, but it requires a large area~\cite{khalil2010loss} and is affected by stray capacities as well.

Besides that, other capacitor structures have been demonstrated: a vacuum-gap~\cite{cicak2010low}, a single-crystal silicon capacitor~\cite{weber2011single} and a parallel-plate capacitor (PPC)~\cite{paik2010reducing, deng2014characterization, cho2013epitaxial}. 
However, few of them can be easily adapted to fabrication of superconducting quantum circuits, such as qubits. Specifically, PPC structures based on a metal-insulator-metal (MIM) stack stand out by their simplicity and reliability. At the same time, PPC is one of the most compact capacitors.

In our paper, we focus on practical aspects of such a capacitor, and we have two main targets. First, to develop a relatively low-loss parallel-plate capacitor with a high capacitance per area, which could be used to construct a compact resonator. Second, to explicitly demonstrate that the size of devices based on this technology is sufficiently less than conventional devices used, and that the performance of those devices is good enough for the superconducting qubit architectures.

In order to estimate the minimal value of $Q_{\mathrm{in}}$ required for the qubit dispersive readout without a Purcel filter, we introduce the criterion: the dispersive shift $\chi$ should be larger than the resonator linewidth $\kappa$, where $\kappa = \omega/Q_{\mathrm{total}} $ and $\chi = g^2/\Delta$ in case of two level-system. If for transmon with multiple levels, the dispersive shift will be much smaller due to higher level dispersive shift~\cite{koch2007charge}. Therefore, $ Q_{\mathrm{total}}\geq  \omega\Delta/g^2$. Using typical experimental numbers the criterion gives $Q_{\mathrm{in}}\geq$ several thousands. In practice, it is possible to resolve the dispersive shift even with larger $\kappa$. Using a Purcell filter, fast dispersive readout with large external decay rate of the readout resonators ($\kappa_{\mathrm{eff}}>5\chi$) has been demonstrated~\cite{sunada2022fast, heinsoo2018rapid}. It gives an estimate for the lowest $Q_{\mathrm{in}} < 1000$. We also aim to reduce the resonator footprint to the same size as a standard qubit, or even more compact. If capacitance and inductance occupy similar areas, and a typical transmon size  is 100~$\mathrm{\mu m^2} \times$ 100~$\mathrm{\mu m^2}$, then $c\geq 10~\mathrm{fF/\mu m^2}$.

For this purpose, one of the reasonable dielectrics for this role is aluminum oxide ($\mathrm{AlO_x}$). Aluminum oxide ($\mathrm{AlO_x}$) is a relatively low-loss dielectric (tan $\delta \sim 3-5 \times 10^{-5}$)~\cite{cho2013epitaxial}. The best $\mathrm{AlO_x}$-based superconducting devices, such as a superconducting qubit, where 1$~\sim~$2~nm thin AlOx layer used as a tunnel barrier, achieve tan $\delta = 5 \times 10^{-7}$~\cite{mamin2021merged}.  In addition, $\mathrm{AlO_x}$ has a large dielectric constant~\cite{kukli2001development} $\varepsilon \sim 10$, which makes $\mathrm{AlO_x}$ a good candidate for a role of dielectric in a PPC capacitor. $\mathrm{AlO_x}$ film can be deposited by various methods: by aluminum evaporation in O$_2$ atmosphere~\cite{antula1969effect, birey1978dielectric}, by thermal or anodic oxidation of a thin aluminum layer~\cite{roy1976electron, dittmer1972electron}, oxidation of a film by plasma~\cite{deng2014characterization}, by RF magnetron sputtering~\cite{rahman1980electrical, rajopadhye1986characterization, brassard2007room, fan2003materials}, by pulsed laser deposition (PLD)~\cite{katiyar2005electrical}, by atomic layer deposition (ALD)~\cite{hu2003mim, klootwijk2008ultrahigh} and by spray pyrolysis~\cite{shamala2004studies}. However, special-purpose equipment is required for these methods in addition to that used for the standard superconducting quantum circuit fabrication. Moreover,  some limiting conditions have to be fulfilled for these methods: extreme temperatures, a break of a vacuum cycle between dielectric and electrode deposition, a need for additional materials, aggressive chemical solutions or plasma conditions. These requirements complicate the fabrication process and can result in the degradation of the performance of fragile elements, such as Josephson junctions, that are widely used in superconducting circuits.

To get around these constraints, we introduce an original method for fabrication of parallel-plate capacitors based on aluminum and aluminum oxide. This method is compatible with superconducting qubit fabrication and requires only an aluminum evaporation system commonly used for Josephson junctions fabrication. The method is robust and reproducible, that confirmed by high fabrication yield for 10 resonator structures and 21 test structures.

\begin{figure}
\includegraphics[width=1\linewidth]{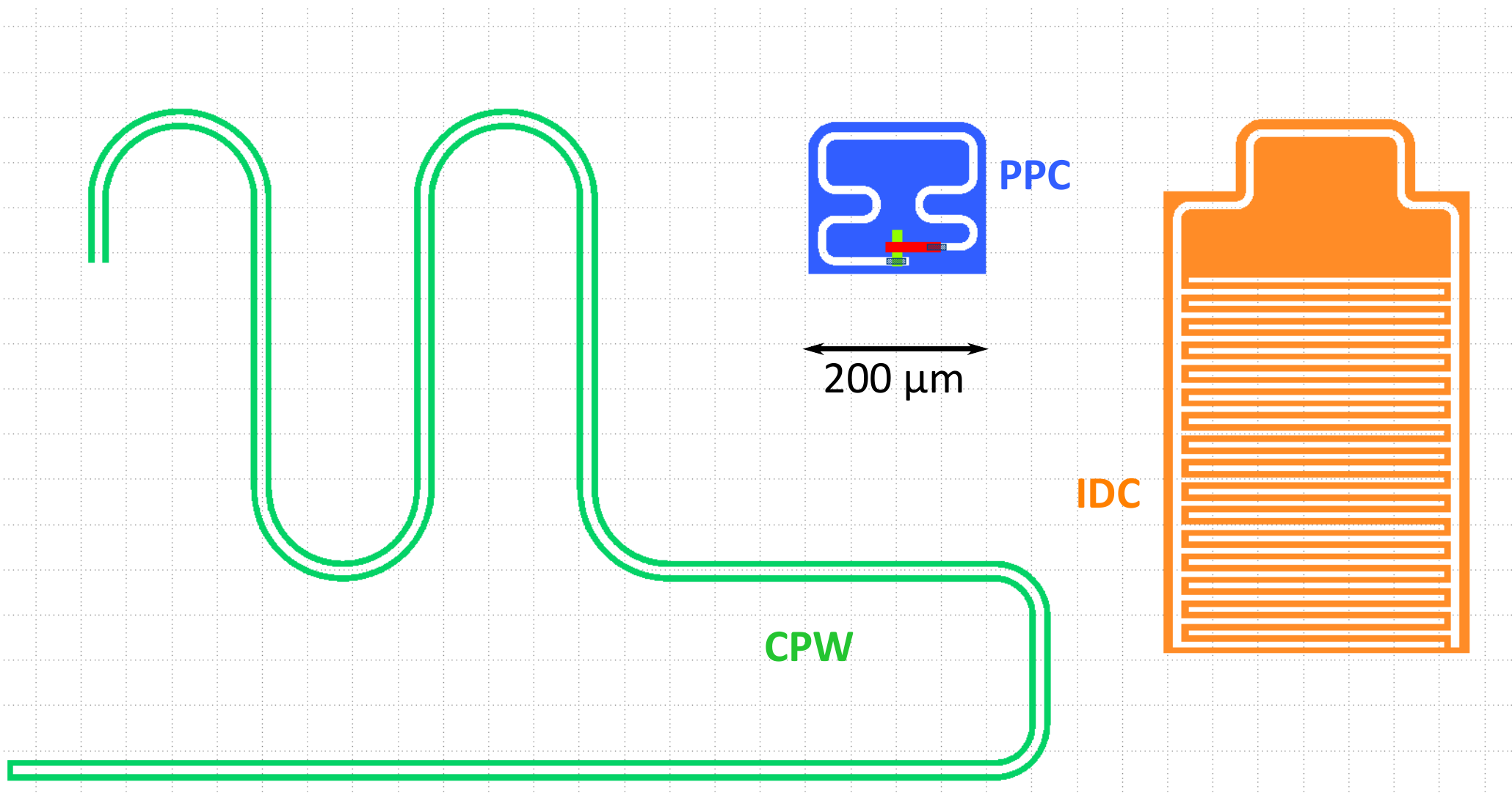}
\caption{Size comparison between three types of superconducting microwave resonators on a silicon substrate: $\lambda/4$~coplanar waveguide, CPW (green); lumped-element based on a parallel-plate capacitor, PPC (blue); and lumped-element based on an interdigital capacitor, IDC (orange). All resonators have a resonance frequency of $f_{r} \sim 7.3$~GHz for illustration purpose.}
\label{1comparison}		
\end{figure}

We characterize dielectric loss by measuring quality factors of lumped-element resonators at microwave frequencies and at low temperatures. The size of such resonators is around 0.04~$\mathrm{mm^2}$, see Fig.~\ref{1comparison}, and internal quality factor $Q_{\mathrm{in}}$ is around $1\times 10^3$ -- $8\times 10^3$ at the single-photon level for resonance frequencies 7--13~GHz.  The capacitance per area is 14~fF/$\mathrm{\mu m^2}$ at microwave frequencies of several gigahertz. 

The method facilitates fast and affordable fabrication recipes and reduces the complexity of the total fabrication. Our results open the road towards the possible applications of such PPC capacitors in compact resonators for qubit readout, impedance matched~\cite{yang2020fabrication} or travelling-wave~\cite{macklin2015near} JPAs, compact Purcell filters~\cite{bronn2015broadband}, kinetic inductance detectors~\cite{beldi2019high} and superconducting metamaterials. This technology paves the way for the miniaturization of essential on-chip components. Also, due to a small value of stray capacitance because of electric field distribution, the PPC capacitors can be used in schemes, where stray capacitance is a crucial hurdle, such as for $0-\pi$ qubit~\cite{gyenis2021experimental}. Furthermore, this technology opens a door for short-wavelength experiments, such as a giant atom-regime~\cite{kannan2020waveguide}.

\section{fabrication process and preliminary test}
The aim is to develop a compact low-loss capacitor fabrication method and to demonstrate that the performance of compact superconducting circuits based on the capacitors is suitable for superconducting qubit operations. The recipe should be easily applicable to superconducting qubit fabrication without high complexity or cost.
 
This section discusses considerations of dielectric thickness. A key element for superconducting circuits, such as qubits, is a Josephson junction.  The state-of-the-art devices use aluminum -- aluminum oxide -- aluminum trilayer.  The typical AlO$_x$ thickness of a Josephson junction is 1--2~nm~\cite{zeng2015direct}. However, for a capacitor, a dielectric layer needs to be thicker to prevent Cooper pair tunneling, to avoid current leakage and a parasitic inductance induced by the Josephson effect.  For example, the thickness of the native aluminum oxide layer is 2--5~nm depending on the aluminum crystallographic orientation~\cite{nguyen2018atomic} for monocrystalline aluminum.  Practically, it is not easy to deposit monocrystalline aluminum during qubit fabrication: aluminum evaporation for Josephson junctions is performed by electron-beam evaporation and hence aluminum is normally amorphous or polycrystalline. For that reason, it is tough to model aluminum oxide growth with uncertain aluminum crystal orientation. 

\begin{figure*}
\includegraphics[width=0.99\linewidth]{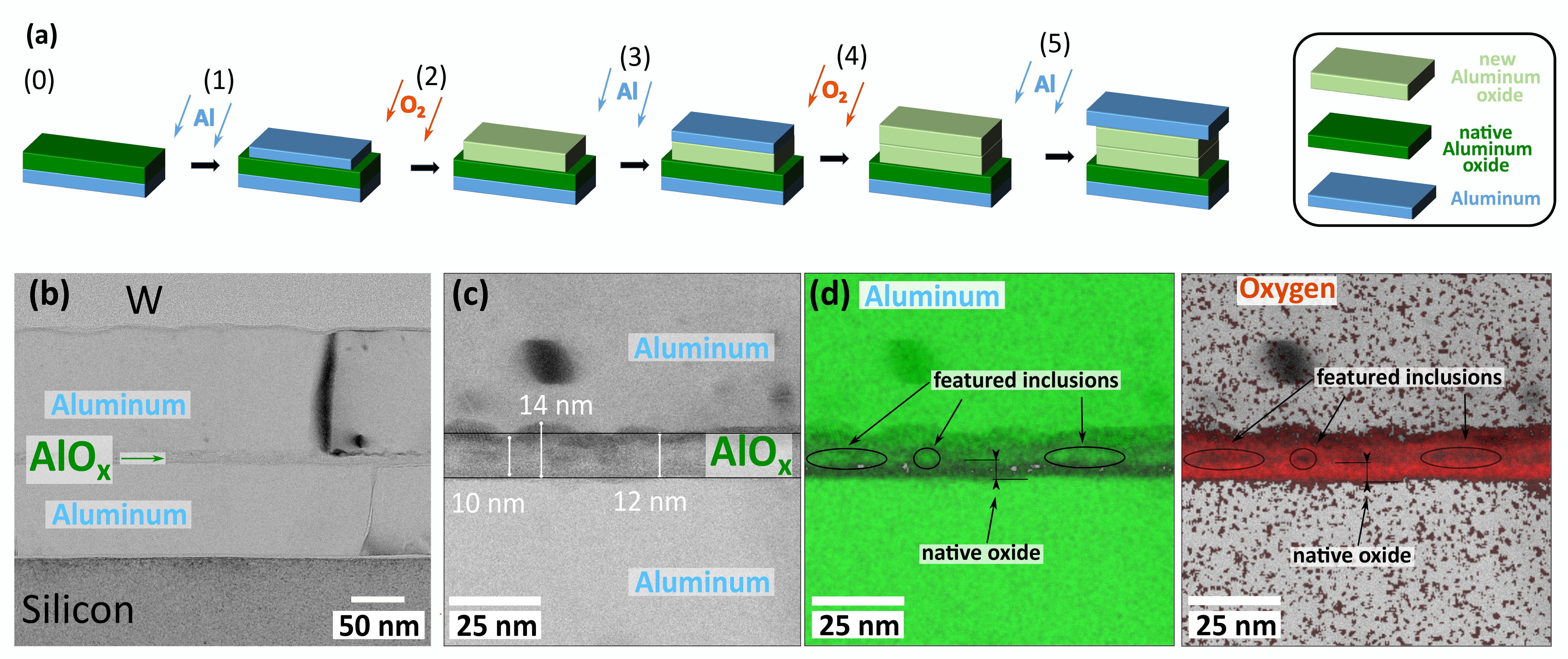}
\caption{(a) A sketch of the PPC  dielectric fabrication process. The fabrication steps are: (0) preparation of a silicon substrate with an already deposited bottom aluminum electrode of a capacitor with native oxide, (1) deposition of the first sacrificed 
aluminum layer, (2) oxidation of the first sacrificed 
aluminum layer, (3) deposition of the second sacrificed 
aluminum layer, (4) oxidation of the
second sacrificed aluminum layer, (5) deposition of the 
second aluminum electrode. (b)-(c) TEM images of the dielectric layer, the average dielectric thickness is $d \sim$ 12~nm. (d)  False-colored TEM image with element mapping of aluminum (blue) and oxygen (orange) using EDX spectroscopy. The brighter a pixel, the more aluminum or oxygen, respectively. Black circles spot the high-$\varepsilon$ features, described in the main text. Black double arrows indicate the native oxide layer.}
\label{2TEM}		
\end{figure*}

First, we prepare a silicon substrate with a aluminum electrode.  The bottom electrode is deposed by an electron-beam evaporator with the thickness of 50~nm, but sputtering method and different thickness can also be used. After liftoff process, the bottom electrode has natural oxide on the surface. We start the description of dielectric formation from this point. To achieve a relatively thick and reproducible aluminum oxide layer, we make a stack of two thin dielectric layers using Plassys MEB550S3, the conventional electron-beam evaporation machine for Josephson junction fabrication. To obtain each dielectric layer, we deposit an aluminum layer of 1~nm and then oxidize it in a pure oxygen atmosphere under normal pressure, as shown in Fig.~\hyperref[2TEM]{\ref*{2TEM}(a)}. The direct thickness measurement of the oxide in Josephson junctions~\cite{zeng2015direct}, that fabricated in the same type of evaporation machine, suggests that 10~minutes in the atmosphere of pure oxygen under 1~atm pressure is enough to grow at least $\sim$1.5~nm of AlO$_x$. Thus, several cycles of 1~nm evaporation + 1~atm oxidation should result in a sufficiently thick oxide layer to avoid current leakage across a capacitor, including the Josephson current and its excess inductance. To check the absence of current leakage, we measure the resistance of the test capacitor structures on a probe station at room temperature. We find that test capacitors with a single additional layer does not have sufficiently reproducible results, thus we focus on testing the PPC capacitors made with two cycles of evaporation and oxidation. We find that the resistance of the capacitors is $R \sim 3\times10^6~\mathrm{\Omega \times \mu m^2}$, while the Josephson junction room temperature resistance is typically $ R_\mathrm{JJ} \sim~10 - 10^3~\mathrm{\Omega \times \mu m^2}$. Hence, we can assume no shunting inductance in our capacitors at cryogenic temperatures, that expressed through room temperature resistance $R$ as  $L =\Phi_0 /(2\pi I_\mathrm{c}(R))$~\cite{schmidt1997physics}, where $I_\mathrm{c}(R)$ is given by Ambegaokar-Baratoff formula~\cite{ambegaokar1963tunneling}.

To investigate the internal structure and  nanoscale property of the dielectric layer we take transmission electron microscopy (TEM) and  energy-dispersive X-ray spectroscopy (EDX) for element mapping analysis. The results are shown in Fig.~\hyperref[2TEM]{\ref*{2TEM}(b)--(c)} and in Fig.~\hyperref[2TEM]{\ref*{2TEM}(d)} respectively. The results spot aluminum (blue) and oxygen (orange) elements. Ones can determine AlO$_\mathrm{x}$ thickness and its morphology, but the boundary between Al and AlO$_\mathrm{x}$ is quite blurry. The AlO$_\mathrm{x}$ thickness is spatially inhomogeneous and varies from $\sim$10~nm to $\sim$14~nm with the average value of $12\pm 2$~nm. The levelized boundary is shown in Fig.~\hyperref[2TEM]{\ref*{2TEM}(c)} by black lines. Also, ones can distinguish the aluminum native oxidation layer as a darker blue layer in Fig.~\hyperref[2TEM]{\ref*{2TEM}(d)} and additionally deposited layers as a brighter blue part. In the middle of the dielectric layer in Fig.~\hyperref[2TEM]{\ref*{2TEM}(d)} there are brighter blue features of the aluminum and darker features in the oxygen figure. These inclusions could be formed as oxygen-depleted aluminum oxide or even incompletely oxidized aluminum. Some of these featured inclusions are highlighted by the black circles. Distribution of these inclusions is inhomogeneous by vertical axis. Due to layer-by-layer fabrication, these features are localized into two horizontal planes, which provide no shorting between the two capacitor plates. Also, it is supported by the results of the dielectric properties test, which is described further.

Since native formation is a process in non-pure oxygen atmosphere, this dielectric layer  may contain more contaminations, such as two-level  (TLS) systems, which can degrade the internal Q factor of the dielectric. Previous research~\cite{melville2020comparison} suggests that the tangent loss of native aluminum oxide can reach up to $<30\times10^{-3}$. In our case, since the  native oxide comprises approximately 25\% of the total dielectric thickness, it can potentially limit the tangent loss. To improve this value, before exposing the freshly deposited metal to the ambient conditions to form  native AlO$_\mathrm{x}$ (after steps 0 and 5, see Fig.\ref{2TEM}), the sample is pre-oxidized  at 15 Torr  for 5 mins to form an initial pre-layer of aluminum oxide with better properties than AlO$_\mathrm{x}$ which  formed immediately at ambient conditions.

\begin{figure}
\includegraphics[width=0.99\linewidth]{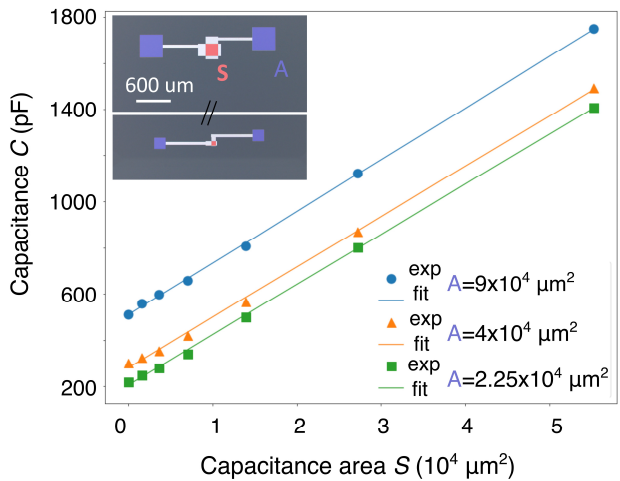}
\caption{Room-temperature capacitance as a function of capacitor area $S$ (overlapping area, red) for a different area of contact pad  $A$ (blue). The data is fit with  $C=c\times S+C_\mathrm{{offset}}$, obtained values of capacitance per area  $c=22.0 \pm 0.5~\mathrm{fF/\mu m^2}$, averaged for different pads size. Inset: false-colored optical image of the test structures.}
\label{3RT_test}
\end{figure}

To test dielectric properties of the capacitors and its reproducibility the capacitors are measured using a capacitance bridge device (Wheatstone bridge). By applying AC at 1~KHz between two probes, the total capacitance value is extracted from the measured a.c. impedance of devices. To make measurements more precise, we choose test capacitance values that are much higher ($\sim$ 100s~pF)  than a minimal value that the bridge device can resolve ($\sim$ 10s~fF), despite in circuit-QED devices much smaller values are usually used ($\sim$ 100s~fF). In the inset of Fig.~\ref{3RT_test}, the test structures are shown. Each test structure is spatially separated by at least 1~mm to avoid parasitic capacitance to ground $C_\mathrm{g}$ and to neighbour structures. To obtain capacitance per area value $c$ we vary overlapping capacitor area $S$. To account for measured offsets due to stray capacitance $C_\mathrm{offset}$, we vary the size of contact pads $A$. The results are shown in Fig.~\ref{3RT_test}.

To determine the capacitance per area $c$ the measured data are fitted to a model of a parallel-plate capacitor $C = c \times S + C_\mathrm{{offset}}$, where $C$ is the total capacitance value, $S$ is the capacitor overlapping area, $C_\mathrm{{offset}}$  is the capacitance value offset, which originates from various sources, such as contact pads, micro-probes, cables etc. Capacitance to ground is negligible there, since the test structures are separated by 1~mm. The obtained value is $c = 22\pm 0.5~\mathrm{fF/\mu m^2}$ at 1~kHz and at 300~K, averaged for three sets of test structures with different contact pad size. The negligible fitting error and high yield demonstrates that the oxidation condition is robust and reliable. 

The dielectric constant of AlO$_\mathrm{x}$ can be found as $\varepsilon=d\times c/\varepsilon_0=30\pm5$ for 1~kHz, where $c$ capacitance per area, $d$ is the thickness of the dielectric layer and the vacuum permittivity is $\varepsilon_0 = 8.85 \times 10^{-12}$~F/m.
 
\section{High-frequency performance at low temperature}

\begin{figure*}
\includegraphics[width=1\linewidth]{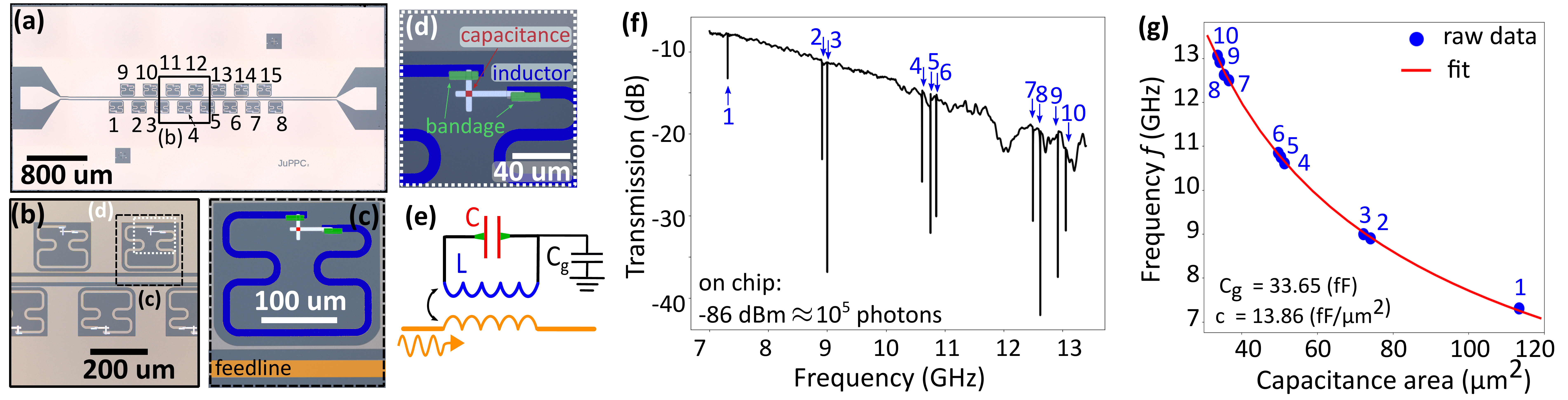}
\caption{(a)--(d) False-colored micrograph of the chip. (e) The equivalent electric scheme, where  $L$ is resonator inductance, $C_\mathrm{g}$ is a capacitance offset. (f) Transmission measurement results for resonance response at high power (-86~dBm on chip), arrows highlight resonance peaks in different groups. (g) Resonance frequency dependence on the capacitor area. The fit function is $f = 1/2\pi \sqrt{L({C_g} + cS)}$, where $c$ is a capacitance per area, $S$ is a capacitor area.}
\label{4PPC}		
\end{figure*}

\begin{table*}
\caption{\label{tab:resdesign}PPC resonator parameters. Measured frequency $f$; fabricated capacitance $C = S \times c$, where $S$ is measured capacitor area  and $c$ is capacitor per area; fabricated square-root of area $\sqrt{S}$, measured  with an optic microscope; $Q_{\mathrm{in}}$: the internal quality factors of the resonator for different applied power on a chip; tangent loss tan~$\delta$.}
\begin{ruledtabular}
\begin{tabular}{ccccccccccc}
         Number of a resonator  & 1 & 2 & 3 &  4 &  5 & 6 & 7 & 8 & 9  & 10 \\
        \hline
        Measured frequency $f$ (GHz) &  7.30  & 8.91 & 9.00 &  10.61 & 10.76 & 10.85 & 12.50 &  12.62 & 12.92 &  13.06  \\

        Fabricated capacitance value $C$ (pF) & 1.56  & 1.02  & 0.99  & 0.70  &  0.69  & 0.68  &
  0.50  & 0.48  & 0.47  & 0.46  \\
        
        Fabricated $\sqrt{\mbox{S}}$ ($\mu$m) & 10.64 & 8.6 & 8.5 &  7.16 & 7.09 & 7.04 & 6.04 &  5.94 & 5.84 & 5.8  \\
        
        average $Q_{\mathrm{ext}}$, ($10^3$) & 9 & 8 & 6 &  6 & 6 & 6 & 6 &  7 & 7 & 9  \\
                
        $Q_{\mathrm{in}}$, $\sim$ $10^5$ photons ($10^3$) & 45.5 & 35.5 & 33.6 &  43.9 & 31.3 & 35.9 & 34.5 &  40.0 & 21.3 & 18.9  \\
         
        $Q_{\mathrm{in}}$, $\sim$ 1 photon ($10^3$) & 4.5 & 2.8 & 1.6 &  3.4 & 0.6 & 2.5 & 3.0 &  8.3 & 0.1 & 2.8  \\
        
        $\mathrm{tan~\delta} = 1/Q_{\mathrm{in}}$, ($10^{-4}$) &   2.22 &  3.57 &   6.25 &   2.94 & 16.6 &   4.00 &   3.33 &   1.20 &  100. &   3.57 \\

\end{tabular}
\end{ruledtabular}
\end{table*}

Superconducting quantum circuits work at low temperatures and microwave frequencies. We examine the capacitance properties under these conditions by testing resonators -- one of the main components in cavity quantum electrodynamics.

The investigated chip is composed of lumped-element resonators inductively coupled to the common feedline. Each resonator consists of a PPC and an inductor made of a wire. The geometric inductance value is kept equal to $L=0.3$~nH and given by finite-element simulations, the capacitance value is varied. The kinetic inductance of the inductor is estimated as 6\% of geometric inductance. The fabrication procedure, described in Chapter II, is used for capacitors formation. Other parts, such as coplanar waveguides, as well as an inductor wire, are made by reactive ion etching of 50~nm niobium film. These parts are galvanically connected in series to the Al-AlOx-Al capacitors via so-called bandages~\cite{dunsworth2017characterization}. Bandages are made by argon milling to remove native aluminum and niobium oxide in the contact areas and the following aluminum deposition for galvanic connection. All structures are photolithographically patterned. The chip is cooled down to about 10~mK in a dilution refrigerator. A vector network analyzer is used to perform transmission measurement $S_{21}$. The chip and the equivalent electric scheme are shown in Fig.~\hyperref[4PPC]{\ref*{4PPC}(a)--(e)}.

First, we test the reproducibility of the capacitance properties at low temperatures and microwave frequencies. To assign each peak to a particular resonator, we separate fifteen resonators into five groups, each containing 1, 2, 3, 4 and 5~resonator peaks. The frequency step inside each group is about 100~MHz and between groups it is about 1~GHz. The resonators are placed next to each other on the same side of a waveguide, as indicated by numbers in Fig.~\hyperref[4PPC]{\ref*{4PPC}(a)}. The measurement results are shown in Fig.~\hyperref[4PPC]{\ref*{4PPC}(f)}. Due to a strong dielectric constant dependence on frequency,  which we will cover in Discussion section, the operation frequency is shifted from 4~--~14~GHz to 7~--~17~GHz. Resonators above 14~GHz are outside of the measurement setup frequency range, hence they are not shown. All other peaks in the measured frequency range are observed and have frequency differences between peaks as is expected from the design, which confirms that the capacitance technology is reliable.


The capacitance per area $c$ is obtained as a parameter of the fitting function $f = \frac{1}{2\pi \sqrt{L(C_\mathrm{g} + cS)}}$, with resonator frequency $f$, inductance $L$, capacitor area $S$ and  $C_\mathrm{g}$ is a capacitance offset, mainly due to capacitance to ground. Unlike the test structure measurement procedure, there is negligibly small stray capacitance caused by contact pads or cables, because of non-direct capacitance measurement. The resonance frequency $f$ as a function of capacitance area $S$ is presented in Fig.~\hyperref[4PPC]{\ref*{4PPC}(g)}. Extracted capacitance per area is $c = 13.86\pm 0.14~\mathrm{fF/\mu m^2}$ and capacitance to the ground is $C_\mathrm{g} = 33.65\pm 6.40~$fF. The dielectric constant is $\varepsilon = 19\pm 3$. Capacitance per area $c$ and $\varepsilon$ are different from the room-temperature test, and in Discussion section we will cover the possible reasons.

\begin{figure*}[!]
\includegraphics[width= 1\linewidth]{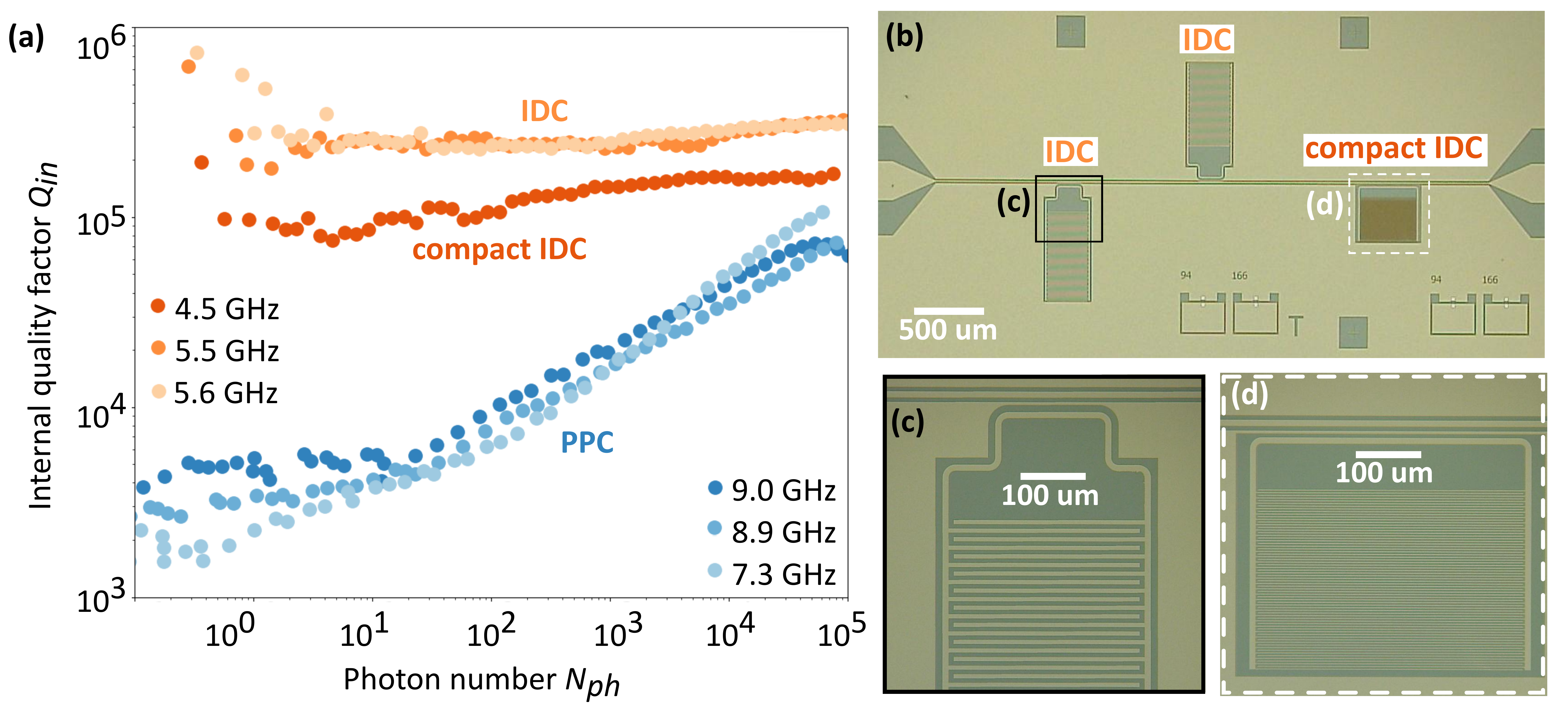}
\caption{(a) Comparison of internal quality factors $Q_{\mathrm{in}}$ versus photon number $N_{\mathrm{ph}}$ in the IDC (orange) and PPC (blue) resonators (see Supplemental Material~\cite{zotova2023compactsupp}, Fig.~2, for plots of $Q_{\mathrm{in}}$ as a function of photon number of all 10 resonators).} (b) Optical micrograph of the sample. (c) Optical micrograph of the IDC resonator. (d) Optical micrograph of the compact IDC resonator with thinner IDC electrode, thus effective capacitor's area is less.
\label{5IDC}		
\end{figure*}

To characterize the magnitude of loss in the PPC resonators, we fit each resonance peak as a function of photon number $N_{\mathrm{ph}}$. It is done using the circle fit technique, implemented in the resonator tool package~\cite{probst2015efficient}. From this fit we extract $Q_{\mathrm{in}}$, $Q_{\mathrm{ext}}$, a resonance frequency $f$, as well as fitting errors. The extracted $Q_{\mathrm{in}}$ of PPC resonators is shown in Fig.~\hyperref[5IDC]{\ref*{5IDC}(a)} as blue dots. $Q_{\mathrm{in}}$ increases when power increases with saturation at low power, which supports the  TLS model of loss~\cite{sage2011study}. The loss is tan~$\delta~\sim~4\times~10^{-4}$ at the low power of a single photon power for the resonance frequency $\sim$ 7--9 GHz.
The tangent loss tan~$\delta$ = $1/Q_{\mathrm{in}}$ that we measure includes not only PPC dielectric loss, but rather all possible sources of internal losses, such as dielectric substrate loss, radiative loss, non-equilibrium quasiparticle loss, vortex loss, and other sources of losses~\cite{mcrae2020materials}. However, the participation of other losses at low power is assumed to be much smaller~\cite{mcrae2020materials} in comparison with PPC dielectric loss, so here we provide a upper bound of the tangent loss of the PPC structure.

The summarized information about resonance frequencies, capacitor areas and capacitance values, obtained quality factors for different photon numbers for each resonator are shown in Table~\ref{tab:resdesign}.

To identify the main source of the loss in PPC resonators, we measure lumped-element resonators with a different type of a capacitor -- an interdigital capacitor (IDC)~\cite{tomonaga2021quasiparticle}--  while the inductor is kept unchanged, bandages are not used. We fit each resonance peak for different photon numbers for each resonator. The extracted internal quality factors are shown in Fig.~\hyperref[5IDC]{\ref*{5IDC}(a)} by orange dots. The measurement results demonstrate a weak dependence of $Q_{\mathrm{in}}$ on power, which illustrates the TLS-loss a small contribution in total loss in IDC structures. 

For PPC resonators, we assume that the bandage part plays an insignificant role in the loss mechanism. Despite the possibility of aluminium oxide remaining in the interface of the bandage,  the bandage provides the galvanic contact. This contact  ensures near-zero voltage drop and, hence, negligible electric field. Thereby, coupling to TLS in the bandage parts and bandage contribution to the loss is assumed to be negligible.

Since the inductor and the entire fabrication process except capacitors are the same for both PPC and IDC resonators, the difference in performance between PPC and IDC resonators is assumed to be related to the capacitor structure.  Due to the different sizes, electric field distributions and dielectrics for IDC and PPC resonators, we do not compare the absolute values of Q-factors, but rather how  Q-factors change with the number of photons. For PPC resonators, the dependence suggests TLS-induced dominant loss, but for IDC is not a TLS mechanism. Therefore, the dependence of Q-factor on the number of photons in Fig.~\hyperref[5IDC]{\ref*{5IDC}(a)} suggests the location of the dominant loss in the PPC resonators within their capacitor.

At low power, the internal quality factors $Q_{\mathrm{in}}$ of the PPC and the IDC resonators differ by two orders of magnitude. In this regime, TLS dielectric loss dominates over other losses in the PPC resonator, coming from TLS inside the PPC. However, for a model, which includes TLS-related and non-TLS dielectric losses, $Q_{\mathrm{in}}$ should have a plateau both at low and high powers. We do not see a second high-power plateau in the internal quality factor of the PPC resonators, but rather a decrease (see Supplemental Material~\cite{zotova2023compactsupp}, Fig.~4). We attribute this drop to other loss mechanisms, such as non-equilibrium quasiparticles. These losses depend on the geometry of the resonators, which is why they manifest themselves in PPC resonators at lower powers than in IDC resonators. 

Also, for the dielectric loss due to TLS, loss tangent gives the frequency dependence $\sim tanh(hf/2kT)$, which is related to the thermal excitation of TLSs at fixed temperature and transition frequency. For GHz range of frequencies, so for all resonators,  and tens of mK corresponds to $tanh(hf/2kT) \sim~$1. There is no significant difference of tangent loss for all PPC resonators.

Also, we compare our fabrication method with previously reported techniques. We focus on dielectric fabrication methods, tan~$\delta$ at single-photon level power, as well as their compactness $c$ and dielectric constant $\varepsilon$. Also, we focus on references, where measurements have been conducted in the GHz frequency range, because dielectric properties are highly frequency dependent.  The summarized results are illustrated in Tab. \ref{tab:c_loss_comparison_type}. 

\begin{table*}
\caption{\label{tab:c_loss_comparison_type} Comparison between this work and previously reported parameters of different dielectric for parallel-plate capacitors. All measurements were performed at low temperatures and at GHz frequency range. The fabrication method: fab. method; the capacitance per area $c$ in $\mathrm{fF/\mu m^2}$ units;  the dielectric constant $\varepsilon$; average tangent loss tan~$\delta$ at  the single-photon level power; thickness of the  dielectric layer $d$ in nanometers are listed.}
\begin{ruledtabular}
\begin{tabular}{ccccccc}
         Ref.& material  & fab. method & $c$ ($\mathrm{fF/\mu m^2}$) & $\varepsilon$ &  tan~$\delta$ &  $d$ (nm) \\
        \hline
        This work & AlO$_x$  & (deposition + oxidation) $\times$ 2 & 14 & 19 &  3$\times$10$^{-4}$  & 12 \\
        \cite{beldi2019high}& AlO$_x$  & ALD & 3.54 & 10 &  1$\times$10$^{-5}$ & 25 \\
        \cite{deng2014characterization}& AlO$_x$  & plasma oxidation & 16 & 9 &  2$\times$10$^{-3}$ & 5 \\
        \cite{mcrae2020dielectric}& AlO$_x$  & sputtering & 2.2 & 12.5 &  9$\times$10$^{-4}$& 50 \\
        \cite{brehm2017transmission}& a-AlO$_x$  & anodic ox. & 1.78 & 10 &  0.4$\times$10$^{-4}$ -- 1.9$\times$10$^{-4}$& 50 \\
        \cite{weides2011coherence}& AlO$_x$  & MBE  & 18 & 12.2 &  6$\times$10$^{-5}$& 6 \\
        \cite{cho2013epitaxial}& AlO$_x$  & MBE  & - & - &  4$\times$10$^{-5}$ -- 6$\times$10$^{-5}$ & 20 -- 40 \\
        \cite{weides2011coherence}& AlO$_x$  & JJ: MBE & 60 & 12.2 &  4$\times$10$^{-5}$& 1.8 \\
       \cite{mamin2021merged}& AlOx  & JJ: oxidation + annealing & 44.3 & 10 &  5$\times$10$^{-7}$ & 2 \\
        \cite{wang2022hexagonal}& hBN  & dry-polymer & 3.3 & 3.76 &  2.5$\times$10$^{-5}$& 10 \\
        \cite{antony2021miniaturizing}& hBN  & dry-polymer & 1.37 & 5.44 &  2.8$\times$10$^{-5}$& 35 \\
        \cite{mcrae2021cryogenic}& GaAs  & MBE & 2.85 & 12.9     &  1.1$\times$10$^{-4}$& 40 \\
        \cite{kim2021enhanced}& AlN  & JJ: MBE & 34.1 & 6.9 &  2.4$\times$10$^{-6}$& 1.8 \\
        \cite{martinis2005decoherence}& a-SiO$_2$  & CVD  & - & - &  5$\times$10$^{-3}$& 300 \\
       \cite{cicak2010low}& SiO$_x$& CVD & -  & -  &  6$\times$10$^{-3}$ & 200\\
        \cite{paik2010reducing}& SiO$_x$& ICP CVD & 0.23  & 6.5  &  1$\times$10$^{-10}$ & 250\\
       \cite{kaiser2010measurement}& SiO$_2$& thermal evap. & -  & -  &  2$\times$10$^{-4}$  -- 4$\times$10$^{-4}$& -\\
        \cite{o2008microwave}& a-Si-H  & sputtering  & - & - &  2$\times$10$^{-5}$& few 100 \\
        \cite{o2008microwave}& SiN$_x$  & sputtering  & - & - &  1$\times$10$^{-4}$& few 100 \\
        \cite{martinis2005decoherence}& a-SiN$_x$  & CVD  & - & - &  2$\times$10$^{-4}$& 300 \\
        \cite{martinis2005decoherence}& a-SiN$_x$  & deposited  by SiH$_4$ and Ni  & - & -- &  2$\times$10$^{-4}$& 9 \\
        \cite{sarabi2016projected}& SiN$_x$ & PE CVD & -  & - &  7.8$\times$10$^{-4}$& 125\\
        \cite{cicak2010low}& SiN$_x$& CVD & -  & -  &  4$\times$10$^{-4}$ & 200\\
        \cite{kaiser2010measurement}& SiN$_x$& PE CVD  & -  & -  &  1$\times$10$^{-4}$  -- 3$\times$10$^{-4}$& -\\
        \cite{antony2021making}& NbSe$_2$  & -  & - & - &  4$\times$10$^{-6}$& 35 \\
        \cite{cicak2010low}& vacuum & suspended & 0.2 & 4.51 &  2$\times$10$^{-5}$ & 200 \\
        \cite{cicak2010low}& a-Si& sputter & -  & -  &  2$\times$10$^{-3}$ & 200\\
       \cite{zhao2020merged}& a-Si  & sputtering  & 17 & 17.5 &  4$\times$10$^{-4}$& 9 \\
        \cite{lecocq2017nonreciprocal}& a-Si& PE CVD & -  & 9  &  1.5$\times$10$^{-4}$ -- 5$\times$10$^{-4}$&  -\\
        \cite{kaiser2010measurement}& Nb$_2$O$_3$& Anodic ox. & -  & 33  &  1$\times$10$^{-3}$  -- 4$\times$10$^{-3}$& -\\
\end{tabular}
\end{ruledtabular}
\end{table*}

Finally, we test the aging degradation of the the structures, contained PPC capacitors. We repeat measurements of an identical sample after 14~months since fabrication. The sample was stored in a cleanroom outside a vacuum desiccator. We find all expected resonators with a frequency difference between samples less than 70~MHz. The capacitance per area is $c = 13.86\pm 0.14~\mathrm{fF/\mu m^2}$ and $C_\mathrm{g} = 34.09\pm 6.2$~fF. At single-photon power the $Q_{\mathrm{in}} \sim 10^3 - 10^4$, it is the same as the measurement of the fresh fabricated sample. The difference between internal quality factors could be caused by the fabrication and measurement setup variation between different samples. The extracted parameters -- resonant frequency  $f$, capacitance per area $c$ and $C_{\mathrm{g}}$ -- have insignificant difference between fresh and aged samples, although we can not estimate the influence of niobium and aluminum degradation separately.

\section{Discussion}
We notice, that capacitor properties are different for different measurement conditions. At the room temperature and at 1~kHz, capacitance per area is $c_{\mathrm{RT}}  \approx 22~\mathrm{fF/\mu m^2}$ and dielectric constant is $\varepsilon_{\mathrm{RT}} \approx 30$. At low-temperature and at 7~--~13~GHz, the values are $c_{\mathrm{LT}} \approx 14~\mathrm{fF/\mu m^2}$ and  $\varepsilon_{\mathrm{LT}} \approx 19$ respectively. Since capacitance area and dielectric thickness are fixed, the difference between capacitance properties should be linked to $\varepsilon$ temperature-frequency behavior. The reported~\cite{seltzman2019precision} temperature dependence of dielectric constant of AlO$_{\mathrm{x}}$ is going to be $d\varepsilon/dT = 1.5 \times 10^{-3}$~/K, which means that the difference between ambient (300~K) and low (10~mK) temperature $\Delta \varepsilon = 0.45~(2\%)$. This value is insufficient to explain the difference we observe between $\varepsilon_{\mathrm{RT}}  \sim 30$ and  $\varepsilon_{\mathrm{LT}} \sim 19$ respectively. 


The dielectric constant $\varepsilon$ is strongly frequency dependent, that has been predicted~\cite{birey1978dielectric} as $\varepsilon (\omega) = \varepsilon_{\infty} + \frac{\varepsilon_\mathrm{s} - \varepsilon_{\infty}}{1+\omega^2\tau^2}$, where $\varepsilon_\mathrm{s}$ and $\varepsilon_{\infty}$ are static and high-frequency dielectric constants, $\tau$ is the characteristic energy relaxation time responsible for dissipation. The mechanism of the frequency dependence of a dielectric constant is a dielectric relaxation response of dipoles to an alternating external electric field, so-called Debye relaxation~\cite{birey1978dielectric}. Previous studies has been reported~\cite{ravindran2006permittivity, suarez2020characterization, dugu2016si, birey1978dielectric, singh2003dielectric} that the dielectric constant $\varepsilon$ of AlO$_{\mathrm{x}}$ changes by 25$\%$--50$\%$ from 1~kHz to 1~MHz. There is expected decreasing trend of dielectric constant $\varepsilon$ value, so we conclude that the difference between capacitance per area for different measurements can be explained by frequency dependence of the dielectric constant $\varepsilon$. However, to better understand it, a further comprehensive study is needed.

It should be noted, that there is a large difference between dielectric constant $\varepsilon$ of our AlO$_{\mathrm{x}}$ and a typical value of $\varepsilon$ for AlO$_{\mathrm{x}}$.

The typical values of $\varepsilon$ for pure AlO$_{\mathrm{x}}$ are in the range  $\varepsilon \sim$ 8-12~\cite{ravindran2006permittivity, kukli2001development, deng2014characterization, beldi2019high, mcrae2020dielectric,brehm2017transmission, weides2011coherence} depending on fabrication method.  For pure Al$_2$O$_3$  $\varepsilon$-frequency dependence is weak compared to dielectrics with conducting inclusions~\cite{ravindran2006permittivity}. 

The difference could be explained by high-$\varepsilon$ featured inclusions embedded into the dielectric layer, as shown in the TEM Fig.~\hyperref[2TEM]{\ref*{2TEM}(d)} above. For example,  aluminum oxide dielectric constant $\varepsilon$ could be doubled by embedding conducting ($\varepsilon \to \infty$) nanoparticles~\cite{ravindran2006permittivity}.  In our case, the aluminum oxidized layer could form oxygen-depleted aluminum oxide or even incompletely oxidized aluminum inclusions, a sign of that we can see in the element mapping as the inhomogeneous vertical color contrast areas,  see Fig.~\hyperref[2TEM]{\ref*{2TEM}(d)}, marked with black circles. 

Future research could be focused on improving capacitance properties, such as compactness or internal losses. To make the capacitance per area  $c = \varepsilon \varepsilon_0/d$ higher one can increase dielectric constant $\varepsilon$ and decrease the dielectric thickness $d$. On the other hand, if performance is as important as compactness,  one should not make the dielectric in a capacitor too thin. It has been shown~\cite{gao2008semiempirical} the TLS noise for parallel-plate capacitance, that induces resonator dephasing, is $S_\mathrm{TLS} = 1/\varepsilon^2 EV$, where $V=Sd$ is a volume of the capacitor dielectric, $S$ is a capacitor area and $d$ is a dielectric thickness. To mitigate dephasing mechanism, we need to increase dielectric constant $\varepsilon$ and make the capacitance volume $V$ as large as possible.  Also, to reduce the dielectric tangent loss tan~$\delta$ it is worth~\cite{martinis2005decoherence} to increase dielectric constant $\varepsilon$ and dielectric thickness $d$, and decrease area $S$. Overall, the optimal strategy  could be increasing dielectric constant $\varepsilon$, and optimizing thickness, which is defined by number of evaporation cycles, and should be determined by particular experiments and their requirements.

To increase the dielectric constant $\varepsilon$ there are several possible strategies: evaporate a thinner layer of aluminum and oxidize it for a shorter time, to embed silver~\cite{ravindran2006permittivity} or other conducting nanoparticles~\cite{kukli2001development, wang2020equivalent}, mixing different dielectric materials~\cite{kukli2001development, siddique2018electric} or deposit it at high temperature (800~C)~\cite{cibert2008properties}. However, embedding conducting nanoparticles could not increase dielectric constant too much. The reason is that if the amount of conductive nanoparticles is significant enough to affect the capacitance, a short circuit can occur with a high probability.

Also, the aluminum films without grains make the dielectric thickness between electrodes more homogeneous~\cite{nik2016correlation}. Future research could be focused on the homogeneity of the films. One of the possible solutions is to vary metal deposition rate and choose electrode thickness carefully~\cite{fritz2019optimization}.

The longer oxidation time enriches the oxide with oxygen and decreases dielectric constant~\cite{singh2003dielectric, ravindran2006permittivity}, as well as capacitance per area value. At the same time, the dissipation factor also could fall, as it has been observed for dielectric with different amount of aluminum inclusions~\cite{singh2003dielectric}  in AlO$_{\mathrm{x}}$ and for AlO$_{\mathrm{x}}$ embedded with silver nanoparticles~\cite{ravindran2006permittivity}. In addition, annealing the capacitor or optimizing the oxidation pressure/time balance can also improve dielectric quality~\cite{mamin2021merged}.

\section{Conclusion}
In summary, this paper shows the method to make a relatively thin 12~nm-thick dielectric layer using only standard equipment for Josephson junction fabrication. We demonstrate room-temperature capacitance measurements and extract capacitance per area  $c~=~22\pm~0.5~\mathrm{fF/\mu m^2}$ at 1~kHz. Also, we test the structures at  cryogenic temperature (10~mK) by making lumped-element resonators.  Resonators $Q_{\mathrm{in}}$ is $1\times 10^3 - 8\times 10^3$ at single-photon level and $Q_{\mathrm{in}}$ is $2\times 10^4 - 4.5\times 10^4$ at $10^5$ photon level. At cryogenic temperatures and GHz frequencies, capacitance per area is $c = 13.86\pm~0.14~\mathrm{fF/\mu m^2}$. The difference between capacitance per area for different measurements could be explained by dielectric constant $\varepsilon$ dependence on frequency, which have been observed~\cite{ravindran2006permittivity, suarez2020characterization, dugu2016si}, and temperature influence assumed to be small~\cite{seltzman2019precision}. The reason the extracted dielectric constant ($\varepsilon_{\mathrm{RT}} = 30 \pm 5$ and $\varepsilon_{\mathrm{LT}} = 19 \pm 3$) is higher than the conventional value $\varepsilon \sim 10$ could be explained by high-$\varepsilon$ inclusion (oxygen-depleted aluminum oxide or even incompletely oxidized aluminum)~\cite{singh2003dielectric, ravindran2006permittivity}. After 14~months of storage circuit parameters with PPC changed less than 1$\%$. 

The demonstrated technology opens the opportunity to make devices with large capacitance values, such as Josephson parametric amplifiers and oscillators, superconducting metamaterials, as well as to shrink the size of the common components of superconducting quantum circuits, such as Purcell filters. Also, compact lumped-element resonators could be used for qubit readout experiments and help with the scaling of many qubit architectures. Finally, the technology can be helpful for experiments with high sensitivity to undesired stray capacitance, such as a $0-\pi$ qubit.

\begin{acknowledgments}
J. Zotova is grateful to Gleb Fedorov, Daria Kalacheva and Ilya Besedin for insightful discussions. This work was funded by RIKEN IPA Program, CREST(JST)  (Grant No. JPMJCR1676), JST [Moonshot R and D][Grant Number JPMJMS2067]  and a project JPNP16007,  commissioned by the New Energy and Industrial Technology Development Organization (NEDO), Japan. The authors are grateful to Russian Science Foundation Project No. 21-42-00025 for financial support.
\end{acknowledgments}

\nocite{*}
\bibliography{apssamp.bib}

\end{document}